\documentclass[10pt,conference]{IEEEtran}
\usepackage{amsmath}
\usepackage{cite}
\usepackage{graphicx}
\usepackage{amsfonts}
\usepackage{latexsym}
\usepackage{subfigure}
\usepackage{algorithm}
\usepackage{algorithmic}
\usepackage{color}
\usepackage{times}
\usepackage{epsfig, graphics}
\usepackage{mathtools}
\usepackage{bm}
\usepackage{wrapfig,lipsum, amsmath, authblk}
\DeclareMathOperator*{\argmax}{arg\,max}
\begin{document}

\newtheorem{theorem}{Theorem}
\newtheorem{lemma}{Lemma}
\newtheorem{conjecture}{Conjecture}
\newtheorem{corollary}{Corollary}
\newtheorem{definition}{Definition}
\newtheorem{scheme}{Scheme}
\newcommand{\rev}[1]{{\color{black}#1}} 
\IEEEoverridecommandlockouts

\newcommand{\final}[1]{{\color{red}#1}} 

\title{Securing NextG Systems against Poisoning Attacks on Federated Learning: A Game-Theoretic Solution\thanks{Authors acknowledge support from the Commonwealth Cyber Initiative (CCI) and the Amazon-Virginia Tech Initiative for Efficient and Robust Machine Learning.}}

\author[1]{Yalin E. Sagduyu}
\author[1]{Tugba Erpek}
\author[2]{Yi Shi}

\affil[1]{\normalsize  Virginia Tech, National Security Institute, Arlington, VA, USA}

\affil[2]{\normalsize  Virginia Tech, Commonwealth Cyber Initiative, Arlington, VA, USA}

\maketitle

\begin{abstract}
This paper studies the poisoning attack and defense interactions in a federated learning (FL) system, specifically in the context of wireless signal classification using deep learning for next-generation (NextG) communications. FL collectively trains a global model without the need for clients to exchange their data samples. By leveraging geographically dispersed clients, the trained global model can be used for incumbent user identification, facilitating spectrum sharing. However, in this distributed learning system, the presence of malicious clients introduces the risk of poisoning the training data to manipulate the global model through falsified local model exchanges. To address this challenge, a proactive defense mechanism is employed in this paper to make informed decisions regarding the admission or rejection of clients participating in FL systems. Consequently, the attack-defense interactions are modeled as a game, centered around the underlying admission and poisoning decisions. First, performance bounds are established, encompassing the best and worst strategies for attackers and defenders. Subsequently, the attack and defense utilities are characterized within the Nash equilibrium, where no player can unilaterally improve its performance given the fixed strategies of others. The results offer insights into novel operational modes that safeguard FL systems against poisoning attacks by quantifying the performance of both attacks and defenses in the context of NextG communications. 
\end{abstract}
\section{Introduction} \label{sec:intro}
\emph{Next-generation (NextG)} communications are anticipated to revolutionize various applications such as smart warehouses, vehicle-to-vehicle networks, and virtual and augmented reality, offering improved throughput, low latency, and enhanced quality of experience (QoE). However, simplistic analytical models based on expert knowledge struggle to capture the intricate nature of waveforms, channels, and resources in NextG communications. To address this challenge, \emph{machine learning} (ML) emerges as a promising approach, enabling learning and adaptation to dynamic spectrum conditions. In particular, \emph{deep learning} (DL) techniques exhibit exceptional capabilities in capturing the high-dimensional data characteristics of wireless communications, surpassing traditional ML methods for detection, classification, and prediction tasks \cite{ErpekBC2019}.

DL-based wireless signal classifiers are effective for incumbent user detection, which is crucial for enabling dynamic spectrum sharing in NextG communication systems with incumbent users. One pertinent use case in the tactical domain involves spectrum co-existence between NextG communication systems and tactical radars, such as in the Citizens Broadband Radio Service (CBRS) band. DL also finds applications in physical layer authentication of user equipment in NextG communications. To address the challenges of data collection from diverse edge devices, \emph{federated learning} (FL) offers a solution by enabling a collaborative system where clients train a single ML/DL model without sharing their individual training datasets. Instead, they share only the model parameters with the server (e.g., base station), which processes the information and  disseminates a global model. This approach not only ensures privacy but also reduces communication load, as the trained models are typically much smaller than the original training data. Moreover, edge devices with limited resources can still participate in collaborative training by using their own capabilities. These capabilities make FL ideal for the collaborative training with multiple sensors in a variety of applications including navigation, autonomy, environment monitoring, and communications systems. To that end, the use of FL over a wireless network should take into account the unique characteristics of wireless communications, ensuring effective training \cite{yang2022federated, niknam2020federated, chen2021distributed, chen2020joint, liu2020secure, shi2022federated,9797923}. 

As DL becomes a core part of NextG systems, there is an increasing concern about the vulnerability of DL to various exploits, attacks, and non-cooperative behaviors. On one hand, clients may be selfish to participate in FL \cite{donahue2021model, sagduyu2022free} by receiving the global model but not returning their trained local models. FL may be also subject to conventional jamming attacks to disrupt communications (model exchanges) between the clients and the server \cite{shi2022launch, shi2023jamming}. In addition, \emph{adversarial machine learning} (AML) equips smart adversaries with novel means to tamper with the training and/or test inputs to DL algorithms for NextG communications. 

The inherent openness and shared nature of the wireless medium make NextG communication systems highly susceptible to threats from adversaries such as eavesdroppers and jammers. These malicious actors have the capability to observe and manipulate the training and test inputs transmitted over the air. Consequently, there is a growing need to understand the attack surface concerning AML attacks targeting the data and control planes of wireless communications \cite{ErpekTCCN2019,adesina2022adversarial, sagduyu2021adversarial, liu2021adversarial}. 

The utilization of distributed clients in FL systems introduces a susceptibility to \emph{poisoning attacks}, which can target multiple clients. In such attacks, a malicious participant injects false or deceptive data into the model, leading to biased or inaccurate results. This compromise can have adverse effects on the global model in FL, causing incorrect decisions to occur.  The vulnerabilities of wireless systems to poisoning attacks have been studied for spectrum access \cite{sagduyu2019adversarial}, cooperative spectrum sensing \cite{luo2020attackers}, and internet of things  \cite{sagduyu2019iot}. 
 
In this paper, we adopt a proactive \emph{defense} approach that focuses on selecting clients for admission into the FL process. The effectiveness of this defense  lies in potentially rejecting the poisoned clients while admitting those that are unpoisoned. By increasing the number of unpoisoned clients admitted and minimizing the inclusion of poisoned clients, the accuracy of the system improves. In order to enhance resilience, it is essential to effectively address the conflicting interactions between the attack and the defense in an analytical framework. \emph{Game theory} provides a mathematical foundation for quantifying the payoffs associated with these interactions, making it a suitable tool for modeling both conventional attacks, such as jamming \cite{sagduyu2011jamming, garnaev2020jamming}, and AML attacks  \cite{sagduyu2022adversarial}. In this paper, we formalize the interactions between the attack and the defense as a \emph{non-cooperative game} and quantify the vulnerability and resilience of FL for NextG signal classification with respect to poisoning attacks. 

The game involves the selection of actions by the attack, namely which clients to poison, and by the defense, namely which clients to admit for FL. The defense aims to maximize the payoff referred to as the utility of resilient operation, which is based on the classification accuracy while considering the cost associated with admitting clients to FL, including computational and communication expenses. Meanwhile, the attack seeks to maximize another payoff, which is the classification error minus the cost incurred when launching the attack, encompassing activities such as eavesdropping, computing, and jamming. We identify the \emph{Nash Equilibrium} strategies as resilient modes of operation, where neither the attack nor the defense can unilaterally change their strategy to improve their payoff given the fixed strategy of the opponent. For this game, we derive equilibrium payoffs as functions of the classification accuracy and costs for both the attack and the defense.

The rest of the paper is organized as follows. Section II provides an overview of FL for wireless signal classification. Section III delves into the attack and defense mechanisms for FL in NextG communications. Section IV introduces the formulated game solution for the scenario involving two clients participating in FL. Building upon this, Section V extends the game formulation to accommodate an increasing number of clients in FL. Section VI concludes the paper.

\section{Federated Learning for Distributed Spectrum Monitoring} \label{sec:fedlearning}
We employ FL to perform spectrum sensing, where a signal classification model is trained using a multitude of edge devices. Each device possesses access to local spectrum data samples. These devices collect spectrum measurements within their respective environments, characterized by varying channel conditions. The collected data is then utilized to train a local model on each device, which is subsequently transmitted to a central server. The central server undertakes the task of aggregating the local models and generating a global model. This global model is then disseminated back to the edge devices for further collaborative training, enabling the joint detection and classification of wireless signals. The FL process can be summarized as follows:
\begin{enumerate}
\item  The server initiates the process by distributing 
the global model network architecture and a copy of the current model weight $w$ to all  clients.

\item Each client $i$ trains its local model $w_i$ with its own data.

\item The clients transmit their local models to the server.

\item The central server aggregates the weight updates from all the clients and computes a new global model weight $w$ by employing federated averaging \cite{mcmahan2017communication}. This updated model is then communicated back to the clients.

\item The clients repeat steps 2--4 until convergence or a predetermined number of rounds is achieved.
\end{enumerate}

The model trained by FL is for wireless signal classification. We assume that the data transmission employs BPSK or QPSK modulation schemes in the background. A group of spectrum sensors, corresponding to a total of $n$ clients in the FL system, is responsible for collecting I/Q data from different locations. The wireless channel introduces path loss, which is influenced by the distance between the transmitter and receiver, as well as random phase shifts. The client locations are assumed to be randomly distributed, and the receiver experiences random noise with a fixed power. From the I/Q data, a set of 16 phase shifts and 16 powers, resulting in 32 features, is utilized to construct a single sample at each client. The samples are then labeled as either 0 (BPSK) or 1 (QPSK). The server's objective is to train a robust classifier capable of identifying BPSK or QPSK signals that are obtained from diverse locations. 

\section{Attack and Defense for Federated Learning} \label{sec:fedsec}
\begin{figure}
    \centering
    \subfigure[Poisoning attack.]
    {
        \includegraphics[width=0.45\columnwidth]{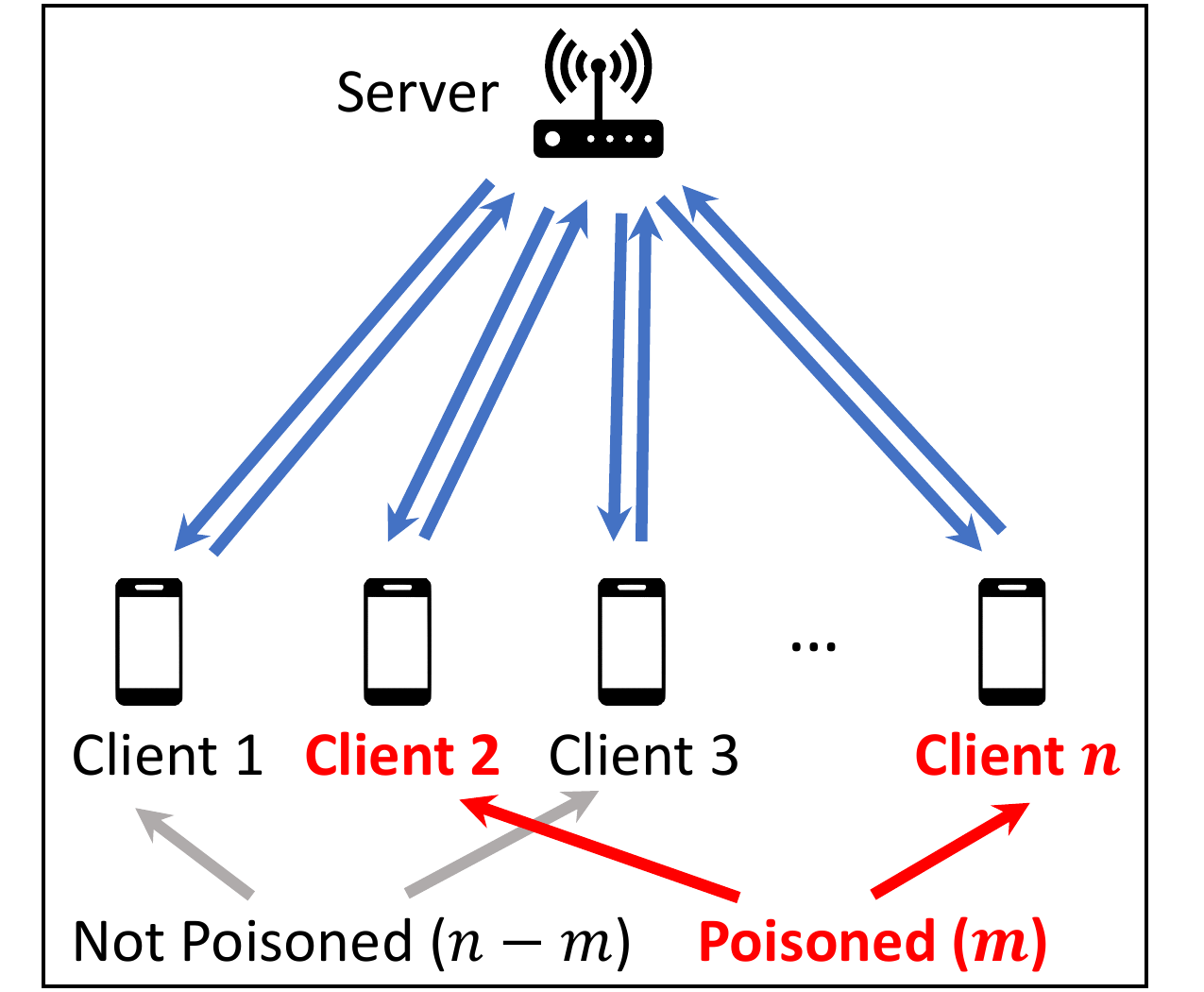}
        \label{fig:FL_A}
    }    
    \subfigure[Defense.]
    {
        \includegraphics[width=0.45\columnwidth]{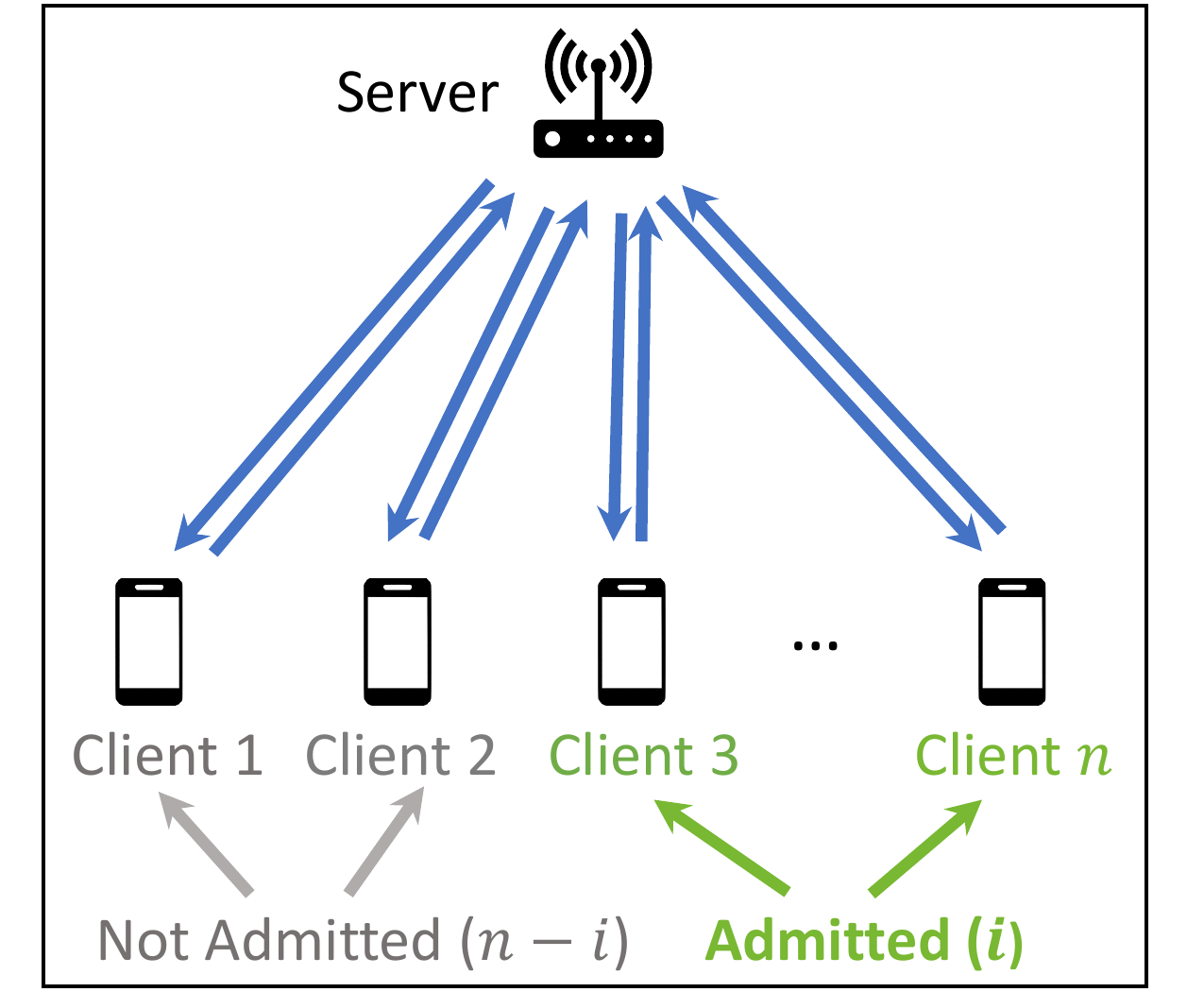}
        \label{fig:FL_D}
    }
    \caption{Poisoning attack and defense.}
    \label{fig:FL_attack_defense}
\end{figure}

We consider the poisoning attack on FL. The attacker's objective is to select specific clients and manipulate their training data samples by altering the labels associated with the data. This results in the generation of distorted model updates that are incorporated into the global training process. The attacker incurs a cost, denoted as $c_A$, for each client that it successfully poisons. To counteract this poisoning attack, we propose a defense mechanism in the form of a client admission policy that promotes resilient operation. Under this defense approach, clients have the autonomy to individually decide whether to participate in the FL process. 
Alternatively, the server can actively select which clients are allowed to participate. The goal is to ensure that only clients with unpoisoned data are admitted, while those with poisoned data are excluded from the FL training process. We consider a cost, denoted as $c_D$, associated with admitting each client. Fig.~\ref{fig:FL_attack_defense} shows the system model for the poisoning attack and defense.

The clients and the server utilize a feedforward neural network (FNN) architecture, characterized by the properties outlined in Table~\ref{table:FFN}. Each client possesses its own local model within the FNN framework. For training, each client contributes 1000 samples. To assess the performance of the global model, an additional 1000 samples, encompassing data from all sensors, are utilized. The accuracy of the FL-trained global model is evaluated after 100 rounds.

\begin{table}
	\caption{FNN properties for wireless signal classifier.}
	\centering
	{\small
		\begin{tabular}{c|c}
		Input size & 32 \\ \hline
		Output layer & size = 2, activation = SoftMax \\ \hline
Dense layers & sizes = $128, 64, 32$, activation = ReLU \\ \hline
Dropout layers &  dropout rate = $0.2$ \\ (after each dense 
 layer) & \\ \hline
Loss function & Crossentropy \\ \hline
Optimizer & RMSprop \\ \hline
Number of parameters & 14,626
		\end{tabular}
	}
	\label{table:FFN}
\end{table}

Clients undergo a selection process to determine their participation in FL, either by being admitted or declined. The effectiveness of this admission policy in maintaining resilience relies on the success in excluding poisoned clients and including unpoisoned clients. Consider total of $n$ clients, $m$ out of them are poisoned (a client is considered poisoned if its training data samples are manipulated), and $i$ out of them are admitted ($n-i$ of them are rejected). Let $k$ denote the number of poisoned clients out of admitted ones. For the worst defense, clients poisoned are the ones admitted such that $k = \min\left(m,i \right)$. For the best defense, clients not admitted ($n-i$ clients) cover poisoned clients ($m$ clients) as much as possible. If $m \leq n-i$, then all poisoned clients are rejected. If not, $m-(n-i)$ poisoned clients are admitted such that $k=\max\left(m-(n-i),0\right)$. Overall, the best and worst defenses would select the admitted clients such that the number of clients poisoned out of admitted clients is given by
\begin{equation}
    k = 
\begin{dcases}
    \max\left(m-(n-i),0\right),& \text{for best defense,}\\
    \min\left(m,i \right),              & \text{for worst defense.}
\end{dcases}
\end{equation}

\begin{figure}[ht]
\centering
\includegraphics[width=0.9\columnwidth]{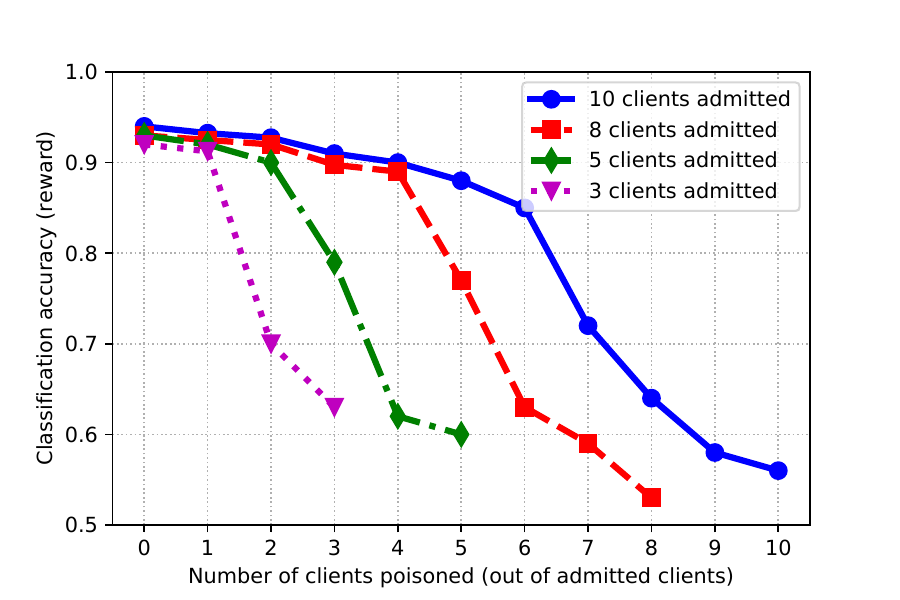}
\caption{Classification accuracy under poisoning attack. }
\label{fig:FL_poisoning}
\end{figure}

Let $U_{k|i}$ represent the classification accuracy, which serves as the reward, in the context of FL when $i$ clients are admitted and $k$ out of these $i$ clients are affected by poisoning. The classification accuracy $U_{k|i}$ for various combinations of $k$ and $i$ is shown in Figure~\ref{fig:FL_poisoning}. As the number of poisoned clients $k$ increases,  $U_{k|i}$ decreases for a given number of admitted clients $i$. Conversely, for a fixed value of $k$, $U_{k|i}$ increases with a larger number of admitted clients $i$ since more unpoisoned clients are included in the FL process. The defense utility is defined as $U_{k|i}- i \: c_D$, where $c_D$ is the cost of admitting one client.

\begin{figure}[h]
\centering
\includegraphics[width=0.9\columnwidth]{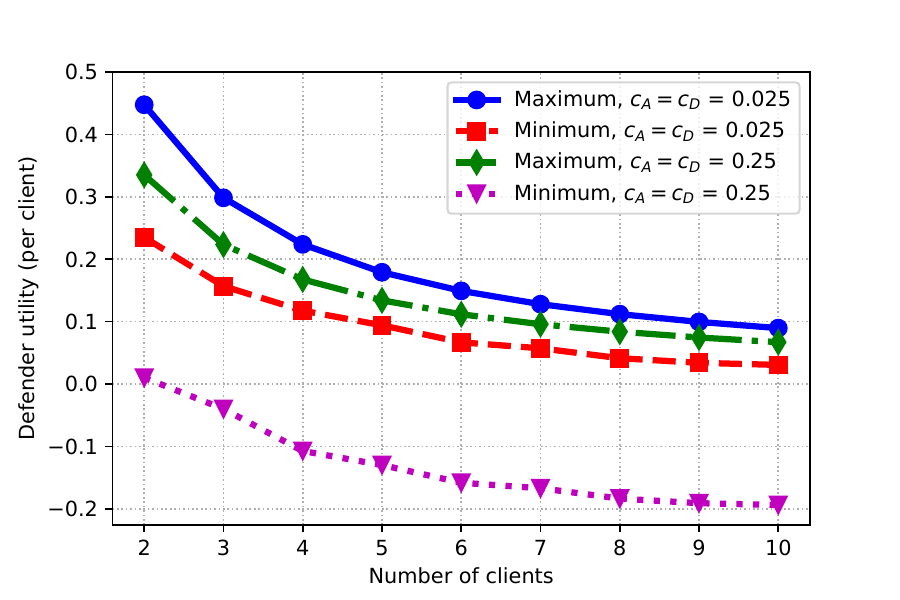}
\caption{The best and worst defense utilities achieved against the poisoning attack.}
\label{fig:new_maxmindefenseutility}
\end{figure}

Fig.~\ref{fig:new_maxmindefenseutility} shows the defense utility as a function of the total number of clients $n$. For a given $n$, the utility is optimized over the selection of the number of admitted and poisoned clients for either the best or worst defense case. These results define the upper and lower limits for the defense utility. In the following sections, we will delve into the analysis of game performance within these established bounds.

\section{Poisoning Attack-Defense Game for Two Clients} \label{sec:gamen2}
We start with two potential clients available for FL. These clients use DL for signal classification. Each client trains its model and sends its parameters to the server. The server processes all local models and broadcasts the global model updates to clients. There is an attacker which selects whether to attack and/or which client(s) to attack. The server decides which clients to accept or reject as a defense method. We formulate the attack and the defense as a non-cooperative game. The defense of excluding some clients' model parameters from the global model update decreases the system performance so it cannot be used all the time. Assume the probability of a server's admitting client $i$ to FL is $q_i$. Let $\mathbf{q}$ denote the vector of admission probabilities. The probability of the attacker's attacking client $i$ is $p_i$. Let $\mathbf{p}$ denote the vector of attack probabilities. 

We assume the server makes independent decisions for the admittance and rejection of each client. Similarly, each client may decide independently on whether to participate in FL, or not. If client $i = 1,2$ participates in FL, its utility is given by  
\begin{eqnarray} 
\label{eq:CND}
\hspace{-0.5cm} && U_{i}^N(q_{-i},\mathbf{p}) =  q_{-i} (p_i p_{-i} (U_{2|2} - c_D) \nonumber \\ \hspace{-0.5cm} 
 && + p_i (1-p_{-i}) (U_{1|2} - c_D)  +   (1-p_i)p_{-i}(U_{1|2} - c_D)  \nonumber \\ \hspace{-0.5cm}  &&   + (1-p_i)(1-p_{-i})(U_{0|2} - c_D))  \nonumber \\ \hspace{-0.5cm}  && + (1-q_{-i}) (p_i(U_{1|1} - c_D) +  (1-p_i)(U_{0|1} - c_D)) ,
 \end{eqnarray}
where $-i$ corresponds to user other than $i$. 

If client $i=1,2$ does not participate in FL, its utility is
\begin{eqnarray}
 \label{eq:CD}
U_{i}^D(q_{-i},\mathbf{p}) & \hspace{-0.2cm} =& \hspace{-0.2cm}  q_{-i} (p_{-i}U_{1|1} + (1-p_{-i})U_{0|1}) \\ \nonumber && \hspace{-0.2cm} + (1-q_{-i}) U_{0|0}
\end{eqnarray}
From (\ref{eq:CND})-(\ref{eq:CD}), client $i$'s average utility (averaged over its strategy $q_i$) is 
\begin{equation}
U_i(\mathbf{q},\mathbf{p}) = q_i U_{i}^N(q_{-i},\mathbf{p}) + (1-q_i)U_{i}^D(q_{-i},\mathbf{p}).
\end{equation}

For given $q_{-i}$ and $\mathbf{p}$, the objective of each client $i=1,2$ is to maximize its average utility at a given time such as
\begin{equation}
\begin{split}
&\max_{q_i: \: 0 \leq q_i \leq 1} \:\: q_i U_{i}^N(q_{-i},\mathbf{p}) + (1-q_i)U_{i}^D(q_{-i},\mathbf{p}) 
\end{split} \label{eq:clientcostmax}
\end{equation}
or, equivalently
\begin{equation}
\begin{split}
&  \min_{q_i: \: 0 \leq q_i \leq 1} \:\: -\left(q_i U_{i}^N(q_{-i},\mathbf{p}) + (1-q_i)U_{i}^D(q_{-i},\mathbf{p}) \right) \\
\end{split} \label{eq:clientcostmin}.
\end{equation}

The solution to the optimization problem in (\ref{eq:clientcostmin}) should satisfy the Karush-Kuhn-Tucker (KKT) conditions, which would ensure a local minimum. The Lagrangian is given by
\begin{eqnarray}
\label{eq:Lagrangian1}
 \hspace{-0.4cm}   \mathcal{L}(q_i; \mu_i^1, \mu_i^2) = && \hspace{-0.6cm} - \left( q_i U_{i}^N(q_{-i},\mathbf{p}) + (1-q_i)U_{i}^D(q_{-i},\mathbf{p}) \right) \\ && \hspace{-0.6cm} - \mu_i^1 q_i + \mu_i^2 (q_i - 1). \nonumber
\end{eqnarray} 

The KKT conditions for $i=1,2$ are given by
\begin{eqnarray} \label{eq:Lagrangian2}
    && \hspace{-0.8cm} -U_{i}^N(q_{-i},\mathbf{p}) + U_{i}^D(q_{-i},\mathbf{p}) - \mu_i^1 + \mu_i^2 = 0, \:\:\: \text{(stationarity)}  \nonumber \\
    && \hspace{-0.8cm} 0 \leq q_i, \:\:\: q_i \leq 1, \:\:\: \text{(primal feasibility)} \nonumber \\
    && \hspace{-0.8cm} \mu_i^1 \geq 0, \mu_i^2 \geq 0, \:\:\: \text{(dual feasibility)} \nonumber  \\
    && \hspace{-0.8cm} \mu_i^1 q_i = 0,  \:\:\:\mu_i^2 (q_i-1) = 0. \:\:\: \text{(complementary slackness)}
\end{eqnarray} 

The optimal choice of $q_i$ should satisfy the KKT conditions. The solution when $\mu_i^1 = \mu_i^2 = 0$ is given by $U_{i}^N(q_{-i},\mathbf{p}) = U_{i}^D(q_{-i},\mathbf{p})$. From (\ref{eq:CND}) and (\ref{eq:CD}), if we assume the symmetric case with $p_i=p$ and $q_i=q$, $i=1,2$, we obtain
\begin{eqnarray}
\label{eq:clientsol2}
&& \hspace{-0.6cm}  q( p^2(U_{2|2} - 2U_{1|2} + U_{0|2})  \nonumber \\ && \hspace{-0.6cm} + p(2U_{1|2} - 2U_{0|2} - U_{1|1} + U_{0|1}) + U_{0|2} - c_D - U_{0|1}) \nonumber \\ && \hspace{-0.6cm} + (1-q)(p(U_{1|1} - U_{0|1}) + U_{0|1} - c_D - U_{0|0})=0     
\end{eqnarray}

The attacker can choose one of the four different options at a given time: attack both of the clients simultaneously, attack only Client 1, attack only Client 2, or attack none of the clients. The utility of the attacker for case 1 when it decides to attack both of the clients simultaneously is given by
\begin{eqnarray}
\label{eq:case1}
 U_{A}^2(q_1,q_2) = && \hspace{-0.6cm} q_1q_2(-U_{2|2}-2c_A)+ \\ && \hspace{-0.6cm} 
 q_1(1-q_2)(-U_{1|1}-2c_A)+ \nonumber \\
 && \hspace{-0.6cm} (1-q_1)q_{2}(-U_{1|1}-2c_A) + \nonumber \\ && \hspace{-0.6cm} (1-q_1)(1-q_2)(-U_{0|0}-2c_A). \nonumber
\end{eqnarray}

The utility of the attacker for case 2 when it decides to attack only Client $1$ is given by
\begin{eqnarray} \label{eq:case2}
  U_{A,1}^1(q_1,q_2) = && \hspace{-0.6cm} q_1 q_2 (-U_{1|2}-c_A)+ \\ && \hspace{-0.6cm} q_1(1-q_2)(-U_{1|1}-c_A) + \nonumber \\ && \hspace{-0.6cm} (1-q_1)q_2(-U_{0|1}-c_A) + \nonumber \\ && \hspace{-0.6cm} (1-q_1)(1-q_2)(-U_{0|0}-c_A). \nonumber
\end{eqnarray}

The utility of the attacker for case 3 when it decides to attack only Client $2$ is given by
\begin{eqnarray}
\label{eq:case3}
  U_{A,2}^1(q_1,q_2) = && \hspace{-0.6cm} q_1 q_2 (-U_{1|2}-c_A)+ \nonumber \\ && \hspace{-0.6cm} q_1(1-q_2)(-U_{0|1}-c_A) + \nonumber \\ && \hspace{-0.6cm} (1-q_1)q_2(-U_{1|1}-c_A) + \\ && \hspace{-0.6cm} (1-q_1)(1-q_2)(-U_{0|0}-c_A). \nonumber 
\end{eqnarray}

The utility of the attacker for case 4 when it decides not to attack is given by
\begin{equation}
 \begin{split}
U_{A}^0(q_1,q_2)= & \:  q_1q_2(-U_{0|2})+q_1(1-q_2)(-U_{0|1}) \\
 & \hspace{-0.2cm} +(1-q_1)q_2(-U_{0|1}) +(1-q_1)(1-q_2)(-U_{0|0}). \label{eq:case4}
 \end{split}
\end{equation}

By averaging over the utilities (\ref{eq:case1})-(\ref{eq:case4}) for four cases, the attacker's average utility is given by 
\begin{eqnarray} \label{eq:avgutility}
U_{A}(\mathbf{q},\mathbf{p}) = && \hspace{-0.6cm} p_1 p_2 U_{A}^2 + p_1 (1-p_{2}) U_{A,1}^1 + p_{2} (1-p_1)U_{A,2}^1 \nonumber \\ && \hspace{-0.6cm} + (1-p_1) (1-p_2)U_{A}^0.
\end{eqnarray}

Given the defender strategy $\mathbf{q}$, the objective of the attacker is to maximize its total utility 
\begin{eqnarray}
\label{eq:attackercostmax}
U_{A}^*(\mathbf{q},\mathbf{p}) = \max_{\mathbf{p}: 0 \leq \mathbf{p} \leq 1} && \hspace{-0.6cm} p_1 p_2 U_{A}^2  p_1 (1-p_{2}) U_{A,1}^1 + \\ && \hspace{-0.7cm} + p_{2} (1-p_1)U_{A,2}^1 + (1-p_1) (1-p_2)U_{A}^0. \nonumber 
\end{eqnarray}

(\ref{eq:attackercostmax}) can be written as a minimization problem as 
\begin{eqnarray}
\label{eq:attackercostmin}
U_{A}^*(\mathbf{q},\mathbf{p}) = \min_{\mathbf{p}: 0 \leq \mathbf{p} \leq 1} && \hspace{-0.6cm} -(p_1 p_2 U_{A}^2 + p_1 (1-p_{2})*U_{A,1}^1  \\ && \hspace{-1cm} + p_{2} (1-p1)U_{A,2}^1 + (1-p_1) (1-p_2)U_{A}^0). \nonumber
\end{eqnarray}

The solution to the optimization problem in  (\ref{eq:attackercostmin}) should also satisfy the Karush-Kuhn-Tucker (KKT) conditions. The Lagrangian is given by
\begin{eqnarray} 
\label{eq:Lagrangian3}
   \hspace{-1cm}  &&  \hspace{-0.5cm} \mathcal{L}(\mathbf{p}; \mu^1, \mu^2, \mu^3, \mu^4) =  -(p_1 p_2 U_{A}^2 + p_1 (1-p_{2}) U_{A,1}^1  \nonumber \\  \hspace{-1cm}&& \hspace{0.75cm} + p_{2} (1-p1)U_{A,2}^1 + (1-p_1) (1-p_2)U_{A}^0)\nonumber \\
     \hspace{-1cm} && \hspace{0.75cm} - \mu^1 p_1 + \mu^2 (p_1-1) - \mu^3 p_2 + \mu^4 (p_2-1). 
\end{eqnarray} 
The KKT conditions are given by
\begin{eqnarray} \label{eq:Lagrangian4}
    && \hspace{-1cm} - p_2 U_{A}^2 - (1-p_{2})*U_{A,1}^1 + p_{2} U_{A,2}^1  \nonumber\\ && \hspace{-0.6cm} + (1-p_2)U_{A}^0 - \mu^1 + \mu^2 = 0,  \text{(stationarity)} \nonumber \\
    && \hspace{-1cm} - p_1 U_{A}^2 + p_1 *U_{A,1}^1 - (1-p_1)U_{A,2}^1 + (1-p_1) 
    U_{A}^0  \nonumber \\ && \hspace{-0.6cm} - \mu^3 + \mu^4 = 0, \text{(stationarity)} \nonumber \\
    && \hspace{-1cm} 0 \leq p_1, 0 \leq p_2, p_1 \leq 1, p_2 \leq 1, \text{(primal feasibility)} \nonumber \\
    && \hspace{-1cm} \mu^1 \geq 0, \mu^2 \geq 0, \mu^3 \geq 0, \mu^4 \geq 0, \text{(dual feasibility)} \nonumber \\
    && \hspace{-1cm} \mu^1 p_1 = 0, \mu^2 (p_1-1) = 0, \text{(complementary slackness)}  \nonumber \\&& \hspace{-1cm} \mu^3 p_2 = 0. \mu^4 (p_2-1) = 0. \text{(complementary slackness)} 
\end{eqnarray} 

The optimal choice of $\mathbf{p}$ should satisfy the KKT conditions. When $\mu^1 = \mu^2 = \mu^3 = \mu^4 = 0$, the solution for the symmetric case of $p_1=p_2=p$ and $q_1=q_2=q$ is given by 
\begin{eqnarray}
\label{eq:attackersol}
&& -pq^2U_{2|2} - q^2U_{1|2} - qU_{1|1} + q^2U_{1|1} - qU_{0|1}  \\ && + q^2U_{0|1} - c_A - q^2U_{0|0} + 2pq^2U_{1|2} 
 + q^2U_{0|2} + 2qU_{0|1}  \nonumber \\ && - 2q^2U_{0|1} + q^2U_{0|0} - pq^2U_{0|2} = 0. \nonumber    
\end{eqnarray}

\begin{figure}[ht]
\vspace{-0.45cm}
\centering
\includegraphics[width=0.9\columnwidth]{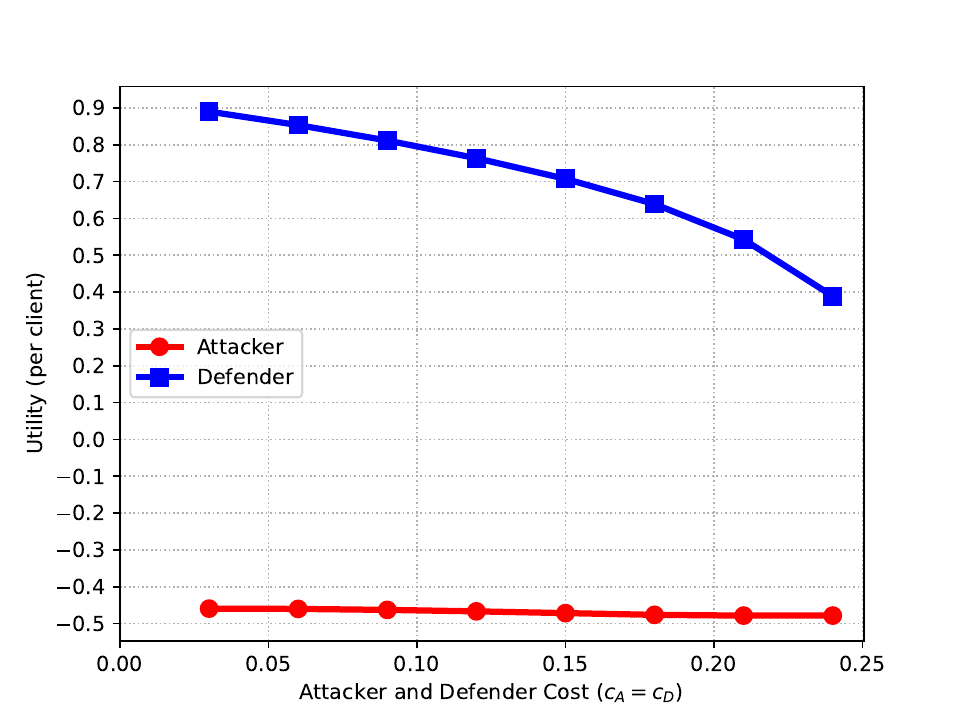}
\caption{Attack and defense utilities per client.}
\label{fig:utility}

\includegraphics[width=0.9\columnwidth]{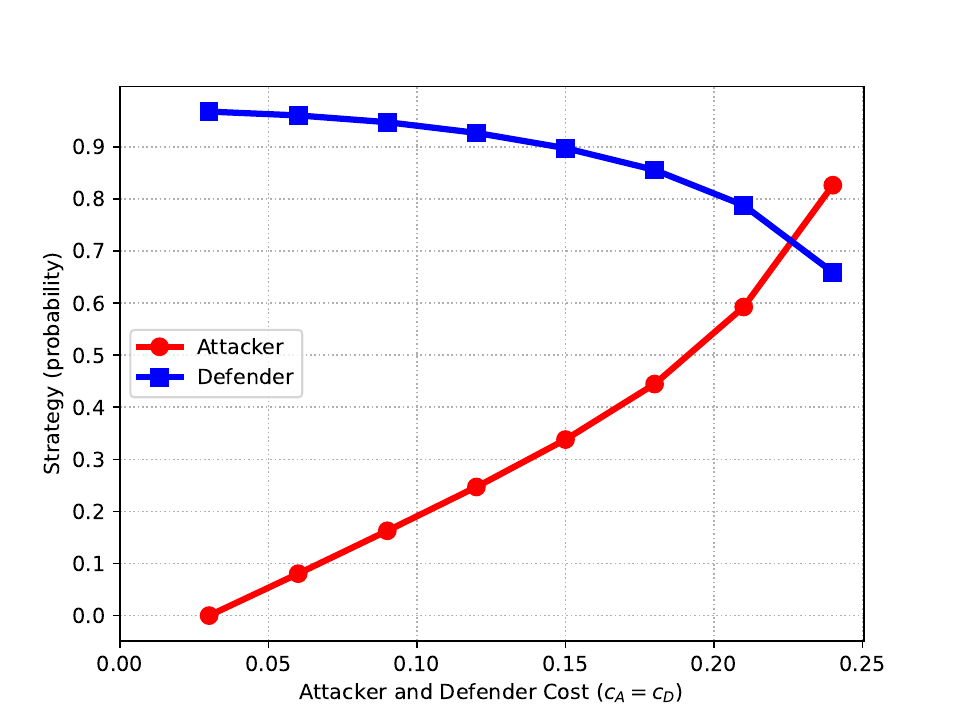}
\caption{Attacker and defender strategies.}
\label{fig:strategy}
\end{figure}

In Nash equilibrium, (\ref{eq:clientsol2}) and (\ref{eq:attackersol}) hold.  Fig.~\ref{fig:utility} and Fig.~\ref{fig:strategy} show the attack and defense utilities and strategies, respectively, as a function of the attack and defense cost. The clients are more declined to participate in FL (i.e., they do not contribute to the global model) with the increasing defense cost. Thus, the defense probability
decreases. This triggers more attacks to increase the impact of the attack. As a result, the attack utility does not decrease with the increasing cost. 

\section{Poisoning Attack-Defense Game for More Than Two Clients} \label{sec:gameng2}
In this section, we increase the number of potential clients to participate in FL. Assume there are $n$ clients in total and each may or may not participate in FL. The defender selects $i$ out of $n$ clients to admit to participate in FL ($n-i$ clients are declined from FL). The attacker randomly selects $m$ clients out of $n$ clients to poison. The utility $U_A(i,m)$ for the attacker consists of two penalties: 
\begin{itemize}
    \item the first penalty is associated with the corresponding FL reward, and
    \item the second penalty is for the cost of poisoning $m$ clients (the cost is $c_A$ for poisoning one client), namely $m\: c_A$ (note that this cost is associated with the data poisoning attack targeting each client separately). 
\end{itemize}  

To compute the first penalty, we need to compute the reward of FL when $i$ out of $n$ clients are admitted and $m$ out of $n$ clients are poisoned (namely, $m \leq n)$. When the attacker selects $m$ clients to poison, it selects those clients out of $n$ clients uniformly randomly. We express this reward in terms of $U_{k|i}$ which is the FL reward when $i$ out of $n$ clients are admitted and $k$ out of $i$ clients are poisoned (namely, $k \leq i$). If $m$ clients are selected out of $n$ clients, some clients are from the set of $i$ admitted clients and the rest are from the set of other $n-i$ clients that are not admitted. 
The contribution from $U_{k|i}$ to the first penalty $U_A(i,m)$ occurs with the probability $\frac{ {i \choose k} {n-i \choose m-k} } { {n \choose m}}$, 
since there are ${n \choose m}$ ways of selecting $m$ poisoned clients from $n$ clients, there are ${i \choose k}$ ways of selecting $k$ of poisoned clients from the $i$ admitted clients and there are ${n-i \choose m-k}$ ways of selecting the rest of ($m-k$) poisoned clients from the clients that are not admitted. Then, the attack utility is given by
\begin{equation} \label{eq:att}
    U_A(i,m) =  \sum_{k=0}^{\min (m,i)}   - \frac{ {i \choose k} {n-i \choose m-k} } { {n \choose m}} U_{k|i} - m \: c_A .
\end{equation}

The utility $U_D(i,m)$ for the defender consists of one reward and one penalty:
\begin{itemize}
    \item the reward is associated with the corresponding FL reward, i.e., the first penalty of the attacker becomes the reward of the defender), when $i$ clients participate in FL and $m$ out of these $i$ clients are poisoned, and
    \item the penalty is for the cost of admitting $i$ clients for FL (the cost is $c_D$ for admitting each client to FL), namely $i\: c_D$ (note that this cost is associated with the information exchange requirement of the server for each client). 
\end{itemize} 
Then, the defense utility is given by 
\begin{equation} \label{eq:def}
    U_D(i,m) =  \sum_{k=0}^{\min (m,i)}    \frac{ {i \choose k} {n-i \choose m-k} } { {n \choose m}} U_{k|i} - i \: c_D.
\end{equation}

We define the best response of the attacker to the defender strategy admitting $i$ clients to FL as 
\begin{equation}
    B_A(i) = \argmax_{0 \leq m \leq n} \: U_A(i,m) . 
\end{equation}

Similarly, we define the best response of the defender to the attacker strategy poisoning $m$ clients as 
\begin{equation} \label{eq:bestm}
    B_D(m) = \argmax_{0 \leq i \leq n} \: U_D(i,m) . 
\end{equation}
From (\ref{eq:att})-(\ref{eq:bestm}), Nash equilibrium strategies $\left(i^*,m^*\right)$ are computed by solving
\begin{equation}
\left(i^*,m^*\right) \in \left(B_D(m^*), B_A(i^*) \right).
\end{equation}

\begin{figure}[t]
\vspace{-0.45cm}
\centering
\includegraphics[width=0.9\columnwidth]{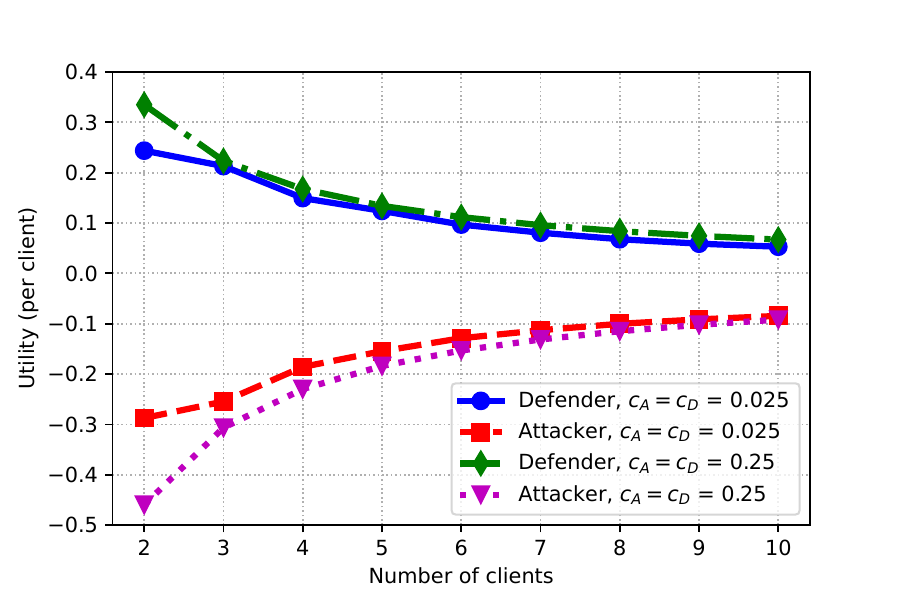}
\caption{Attack and defense utilities in Nash Equilibrium as we increase the number of clients.}
\label{fig:Nash_gen_n}
\end{figure}

Fig.~\ref{fig:Nash_gen_n} shows the attack and defense utilities, $U_A(i^*,m^*)$ and $U_D(i^*,m^*)$, in Nash Equilibrium as a function of the number of clients $n$. As $n$ increases, the defense utility per client decreases as it is easier to poison some of the increasing number of clients and it becomes costly to admit more clients. In response, the attack utility increases. As we increase the attack and defense cost, the attack utility decreases as it becomes more costly to attack. This decrease helps the defender compensate the increase of its own cost and eventually increase its utility for resilient FL.  

\section{Conclusion} \label{sec:conclusion}
We considered the poisoning attack on FL systems for wireless signal classification in NextG communications and analyzed the attack-defense interactions in a game setting to characterize resilient operation modes for FL. In the poisoning attack, training data of any client may be poisoned to mislead the training process for the global model. The defense may select between admitting or rejecting clients with the ultimate goal of eliminating poisoned clients. We formulated a non-cooperative game played between the attacker and the defender to quantify the performance due to the conflicting interests. After deriving the attack and defense performance bounds, we determined the attack and defense strategies and utilities in Nash equilibrium for two clients and extended the analysis to an arbitrary number of clients. This analysis has led to novel resilient operation modes identifying how to protect FL against poisoning attacks in NextG communication systems. 

\bibliographystyle{ieeetr}
\bibliography{references}
\end{document}